\documentclass[preprint,showpacs,floatfix]{revtex4}

\usepackage{graphicx}

\begin{document}
\def\d {\dagger}
\def\e {\epsilon}
\def\Rb {{\bf R}}
\def\Tb{ {\bf T}}
\def\sss{\scriptscriptstyle}
\def\ss{\scriptstyle}
\def\bb {{\bf b}}
\def\E {{{\cal E}}}
\def\ninj#1#2#3#4#5#6#7#8#9{\left\{\negthinspace\begin{array}{ccc}
#1&#2&#3\\#4&#5&#6\\#7&#8&#9\end{array}\right\}}
\def\sixj#1#2#3#4#5#6{\left\{\negthinspace\begin{array}{ccc}
#1&#2&#3\\#4&#5&#6\end{array}\right\}}
\def\M {{{\cal M}}}
\def\sqi{\frac{1}{\sqrt{2}}}
\def\x{\times}
\def\nn{\nonumber }
\def\w {{\omega}}
\def\endauthors{}
\def\authors#1\endauthors{#1}
\def\fot{\frac{1}{2}}
\def\tfot{\frac{3}{2}}
\def\mbs{\mbox{\boldmath$\sigma$}}
\def\mbp{\mbox{\boldmath$\phi$}}
\def\mbpi{\mbox{\boldmath$\pi$}}
\def\mbt{\mbox{\boldmath$\tau$}}
\def\bin#1#2{\left(\negthinspace\begin{array}{c}#1\\#2\end{array}\right)}
\def\binv#1#2{\left\{\negthinspace\begin{array}{l}#1\\#2\end{array}\right.}
\def\rf#1{{(\ref{#1})}}
\def\ov#1#2{\langle #1 | #2  \rangle }
\def\sss{\scriptscriptstyle}
\def\ss{\scriptstyle}
\def\bra#1{\langle #1|}
\def\ket#1{|#1 \rangle}
\def\Ket#1{||#1 \rangle}
\def\Bra#1{\langle #1||}
\def\rb {{\bf r}}
\def\pb {{\bf p}}
\def\Pb{ {\bf P}}
\def\be{\begin{equation}}
\def\ee{\end{equation}}
\def\br{\begin{eqnarray}}
\def\er{\end{eqnarray}}
\def\binv#1#2{\negthinspace\begin{array}{c}#1\\#2\end{array}}
\def\cuav#1#2#3#4{\negthinspace\begin{array}{c}#1\\#2\\#3\\#4\end{array}}
\def\binn#1#2{\left\{\negthinspace\begin{array}{l}#1\\#2\end{array}\right.}
\def\jp{{\sf{j}}_p}
\def\jN{{\sf{j}}_N}
\def\jL{{\sf{j}}_\Lambda}
\def\js{{\sf{j}}}
\def\trinn#1#2#3{\left\{\negthinspace\begin{array}{l}#1\\#2\\#3\end{array}\right.}

\begin{titlepage}
\pagestyle{empty} \baselineskip=21pt
\begin{center}
{\large{\bf Deformation and shell effects in nuclear mass formulas}}
\end{center}
\authors
\centerline{C\'esar Barbero$^{1,2}$, Jorge G. Hirsch$^{3,*}$ and Alejandro Mariano$^{1,2}$}
\vskip -.05in
\centerline{\small \it $^{1}$ Departamento de F\'{\i}sica, Universidad Nacional de La Plata, C.
C. 67, 1900 La Plata, Argentina}
\vskip -.05in
\centerline{\small \it $^{2}$ Instituto de F\'{\i}sica La Plata, CONICET, 1900 La Plata, Argentina}
\vskip -.05in
\centerline{\small \it $^{3}$ Instituto de Ciencias Nucleares, Universidad Nacional Aut\'onoma de M\'exico, 04510 M\'exico D.F., M\'exico}
\endauthors

\bigskip

\centerline{ {\bf Abstract} }
\baselineskip=18pt
\noindent

We analyze the ability of the three different Liquid Drop Mass (LDM) formulas to describe nuclear masses for nuclei in various deformation regions. Separating the $2149$ measured nuclear species in eight sets with similar quadrupole deformations, we show that the masses of prolate deformed nuclei are better described than those of spherical ones. In fact, the prolate deformed nuclei are fitted with an RMS smaller than $750$ keV, while for spherical and semi-magic species the RMS is always larger than $2000$ keV. These results are found to be independent of pairing.

The macroscopic sector of the Duflo-Zuker (DZ) mass model reproduces shell effects, while most of the deformation dependence is lost and the RMS is larger than in any LDM.
Adding to the LDM the microscopically motivated DZ master terms introduces the shell effects, allowing for a significant reduction in the RMS of the fit but still exhibiting a better description of prolate deformed nuclei. The inclusion of shell effects following the Interacting Boson Model's ideas produces similar results.
\bigskip



{$^*$ Email}: {hirsch@nucleares.unam.mx}

\end{titlepage}

\baselineskip=21pt

\bigskip

\section{Introduction}

\bigskip

When nuclear physicists refer to the description of nuclear masses employing the Liquid Drop (LD) model formula, we have a contradictory speech. In fact, sometimes we say: "The liquid-drop energy of a spherical nucleus is described by a Bethe-Weizs\"acker mass formula" (\cite{Wang10} citing \cite{Bet36}) or "This is a crude model that does not explain all the properties of the nucleus, but does explain the spherical shape of most nuclei" \cite{Wik}. However, at the same time, we say:
"The semi-empirical mass formula gives a good approximation for atomic masses and several other effects, but does not explain the appearance of magic numbers." \cite{Wik}, which are usually considered the 'more spherical' nuclei, because 
deformation is associated to the quadrupolar interaction between valence protons and neutrons \cite{Fed79}. 
The usual procedure to find a phenomenological mass formula which reduce the root mean square (RMS) for the actually measured $2149$ nuclear species, is to start with the Liquid Drop Mass (LDM) and add to it corrections due to deformation and shell-effects.

The description of nuclear masses in terms of the LDM paved the way to the basic understanding of nuclear properties, like the saturation of the nuclear force, the existence of pairing and shell effects, and the description of fission and fusion processes \cite{Bohr}. The Q-values of different nuclear reactions, obtained from mass differences, must be accurately known to allow the description of the astrophysical origin of the elements \cite{Rol88}. Accurate theoretical predictions of nuclear masses remain a challenge \cite{Bla06}, sharing the difficulties with other quantum many-body calculations, and complicated by the absence of a full theory of the nuclear interaction.

Decades of work have produced microscopic and macroscopic mass formulas \cite{Lunn03}. At present, the most successful approaches seem to be the microscopic-macroscopic models, like the Finite Range Droplet Model (FRDM) \cite{Mol95}, its improvements \cite{Wan10}, and the realistic Thomas-Fermi ͑(TF͒) models \cite{Mye96,Pom03}, the Skyrme and Gogny Hartee Fock Bogolyubov (HFB) \cite{Gor01,Gor09}, and the Duflo-Zuker (DZ) mass formula \cite{Duf94,Zuk94,Duf95}. They allow for the calculation of masses, charge radii, deformations, and in some cases also fission barriers. They all contain a macroscopic sector which resembles the LDM formula, and include deformation effects.
HFB calculations are now able to fit known nuclear masses with deviations competitive with the microscopic-macroscopic
calculations, while the most precise and robust nuclear mass predictions are given by the DZ model \cite{Lunn03,Men08}, which gives an RMS of $373$ keV.

Efforts for building algebraic nuclear mass formulas inspired in the DZ model success have led to detailed analysis of the microscopic building blocks of this model \cite{Men10}, which suggested new ways to introduce the shell effects through the DZ master terms \cite{Hir10}. Other line of thought connected the Interacting Boson Model (IBM) F-spin with the DZ microscopic terms \cite{Die07,Men08,Die09}, which allows for very good fits of the nuclear binding energies when an additional one-body Hamiltonian with a large number of parameters is employed \cite{Gan11}.

Following these works, 
we perform in this paper an analysis of the interplay between deformations and shell effects for different nuclear mass formulas. We start studying the ability of LDM to describe nuclear masses for nuclei in different deformation regions and explicitly show that the best fit is obtained for deformed nuclei, extending a previous preliminar work \cite{Hir11}.
These results are found to be independent of the pairing term, which could be failing in the vicinity of closed shell nuclei.

Analyzing the macroscopic terms of the DZ mass model we show that shell effects are reproduced and most of the deformation dependence is lost and the RMS is larger than in any LDM.
Adding to the LDM the microscopically motivated DZ master terms introduces the shell effects, allowing for a significant reduction in the RMS of the fit but still exhibiting a better description of prolate deformed nuclei. The inclusion of shell effects through a dependence in the number of valence nucleons produces similar results.

The paper is organized as follows:  the  fits for three LDM are presented in Sect. 2, and the DZ and other microscopic algebraic estimations are discussed in Sect. 3. A meticulous analysis of the master terms is performed in Sect. 4, and a comparison with other estimations based on IBM is given in Sect. 5. Conclusions are drawn in Sect. 6.

\bigskip

\section{The fits for three LDM formulas}

\bigskip

We have selected three LDM formulas to analyze their ability to fit nuclear masses.

\bigskip

\subsection{LDM1}

\bigskip

The first one is an improved version of the LDM formula with modified symmetry and Coulomb terms, built following a consistent treatment of nuclear bulk and surface effects \cite{Die09}. The negative nuclear 
binding
energy is given by
\begin{equation}
E_{LDM1} = -a_v  A + a_s  A^{2/3}
+ S_v \frac {4 T(T+1)}{A(1+yA^{-1/3})}
+ a_c \frac{Z (Z-1)}{(1-\Lambda) A^{1/3}}
- a_p \frac {\Delta}{A^{1/3}},
\label{piet}
\end{equation}
where: i) the pairing interaction is given by $\Delta$ = 2, 1, and 0 for even-even, odd-mass and odd-odd nuclei, respectively; ii) a correction to the radius of the nucleus is included through a modification $\Lambda$ in the Coulomb term, $\Lambda = \frac {N - Z} {6Z(1 + y^{-1}A^{1/3})}$; iii) the symmetry term employs $4T(T + 1)$, with $T=|N-Z|/2$, instead of $(N - Z)^2$ to account for the Wigner energy; and iv) the Coulomb interaction is proportional to  $Z(Z - 1)$ to avoid the Coulomb interaction of a proton with itself.

\bigskip

\subsection{LDM2}

\bigskip

The second LDM formula is a modified version of the Bethe-Weizs\"acker one, which incorporates explicitly isospin effects  \cite{Wang10}:
\begin{equation}
E_{LDM2} = -a_v  A + a_s  A^{2/3}
+ a_{sym} I^2 A
+ a_c \frac{Z (Z-1)}{A^{1/3}} (1- Z^{2/3})
- a_{pair} \frac {\delta_{np}}{A^{1/3}},
\label{wang}
\end{equation}
with isospin asymmetry $I = (N - Z)/A$. The pairing term is taken from \cite{Men10}
\begin{equation}
\delta_{np} = \left\{
\begin{array}{ll}
2 - \mid I \mid &\hbox{ :N and Z even,}\\
\mid I \mid &\hbox{ :N and Z odd,}\\
1 -\mid I \mid &\hbox{ :N even, Z odd, and N$>$Z,}\\
1 - \mid I \mid &\hbox{ :N odd, Z even, and N$<$Z,}\\
1 &\hbox{ :N even, Z odd, and N$<$Z,}\\
1 &\hbox{ :N odd, Z even, and N$>$Z,}
\end{array}
\right.
\end{equation}
and the symmetry energy coefficient, including
an $I$ correction on the conventional surface-symmetry term of LDM
to approximately describe the Wigner effect for heavy nuclei, is written as
\begin{equation}
a_{sym} = c_{sym} \left[ 1 - \frac {\kappa} {A^{1/3}}  + \frac {2 - \mid I \mid} {2 + \mid I \mid A} \right].
\end{equation}

\bigskip

\subsection{LDM3}

\bigskip

The third LDM formula is derived from the liquid drop model with the pairing energies of the Thomas-Fermi model \cite{Roger10}
\begin{equation}
\begin{array}{rl}
E_{LDM3}(N,Z) =& -a_v (1 - k_v I^2) A + a_s  (1 - k_s I^2) A^{2/3} + a_k  (1 - k_k I^2) A^{1/3}
+ \frac 3 5 \frac{e^2 Z^2}{r_0 A^{1/3}}
\\&- f_p Z^2/A  - a_{c,exc} Z^{4/3}/A^{1/3}
+ E_{pair}.
\end{array}
\label{roger}
\end{equation}
It includes: i) a first term representing the volume energy corresponding to the saturated exchange force and infinite nuclear matter, with $I^2 A$ being the asymmetry energy of the Bethe-Weizs\"acker mass formula; ii) a surface energy corresponding to semi-infinite nuclear matter and originated by the deficit of binding energy of the nucleons at the nuclear surface; iii) a curvature energy resulting from non-uniform properties which correct the surface energy and depends on the mean local curvature; iv) the decrease of binding energy due to the Coulomb repulsion; v)
a $Z^2/A$ diffuseness correction to the sharp radius Coulomb energy; vi) a $Z^{4/3}/A^{1/3}$ charge exchange correction term; vii)
the pairing energies $E_{pair}$ of the Thomas-Fermi model (taken from Eq. (A.2) in Ref. \cite{Mye96}).

\bigskip

\subsection{The fits}

\bigskip

The coefficients of the three LD models were selected to minimize the RMS when the predicted binding energies $BE_{\rm th}(N,Z)$
are compared with the experimental ones $BE_{\rm exp}(N,Z)$, reported
in AME03 \cite{AME03}, modified so as to include more realistically the electron
binding energies as explained in Appendix A of Lunney, Pearson and
Thibault~\cite{Lunn03}:
\begin{equation}
{\rm RMS}=\left\{\frac{{\sum\left[BE_{\rm exp}(N,Z)-BE_{\rm
        th}(N,Z)\right]^2}}{N_{nucl}}\right\}^{1/2}.
\end{equation}
The minimization procedure uses the routine Minuit \cite{Minuit}.
\begin{table}[h]
\begin{center}
\caption{The nine groups of nuclei employed in the present study, their range of quadrupole deformation, and their number of nuclei.}
\label{tab1}
\begin{tabular}{c||c|ccccccc|c}
\hline \hline
group & all  &	1	&2	&3	&4	&5	&6	&7	&semi-magic\\
\hline
$e_2$ min & -0.65 & -0.65  & -0.11 &	0.00 &	0.04 & 0.12 & 0.18 & 0.23 &\\
$e_2$ max &	0.65 & -0.11 & 0.00  & 0.04 & 0.12 & 0.18 & 0.23 &  0.65 &\\
$N_{nucl}$  &2149 & 258	& 252  &332	       &272       & 307 & 364  & 364   &185\\
\hline\hline \end{tabular} \end{center}
\end{table}

\begin{figure}[h]
\vspace{-3.5cm}
\includegraphics[width=17.0cm]{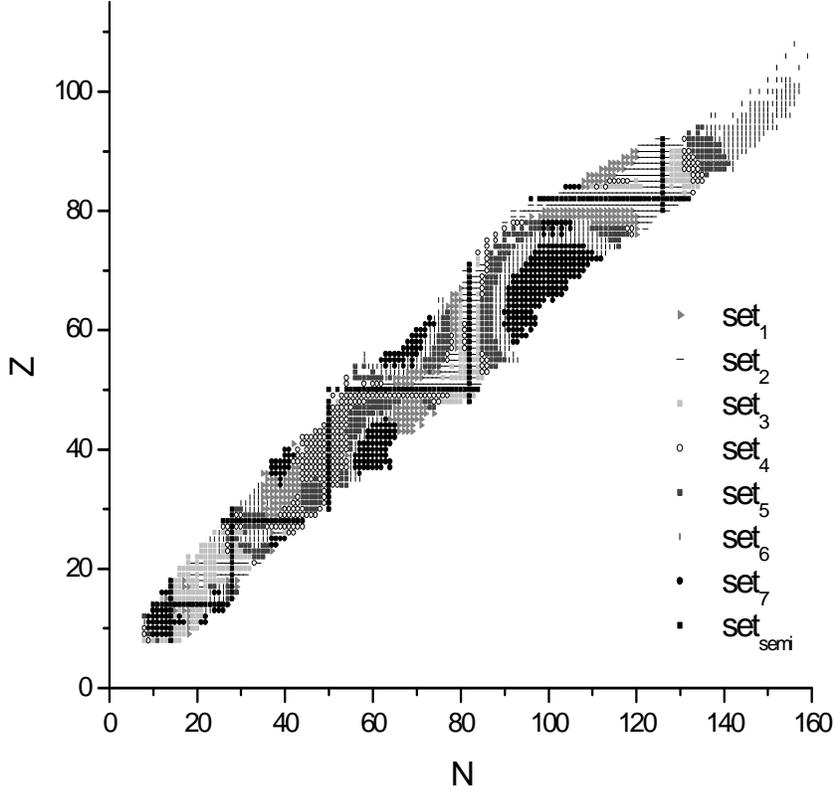}
\vspace{-8cm}
\caption{The seven deformation regions and the semimagic nuclei, shown in the $N$-$Z$ plane.}
\label{fig1}
\end{figure}

The fits were performed separating the nuclei in nine groups:
\begin{itemize}
\item the first one contains all nuclei whose measured masses are reported in AME03 \cite{AME03}, which have $N,\,Z \geq 8$,
\item the next seven groups contain the nuclei whose quadrupole deformations $e_2$, taken form the FRDM \cite{Moll95}, lie in the ranges listed in the second and third row of Table \ref{tab1},
\item the last group contains all semi-magic nuclei, having $Z= 14$, $28$, $50$, $82$ or  $N=   14$, $28$, $50$, $82$ or $126$.
\end{itemize}
$N_{nucl}$ is the number of nuclei in each group, listed in the fourth row of Table \ref{tab1}. We remark that regions $1$ to $7$ contain approximately the same quantity of nuclei.
Notice that group 1 contains most of the oblate nuclei, that the more spherical nuclei belong to groups 2, 3 and 4, and that the more prolate deformed nuclei are included in groups 6 and 7. We show in Fig. \ref{fig1} the different regions in the N-Z plane.  Semimagic nuclei are displayed along straight thick black lines. Around them cluster the nuclei classified as the more spherical, while the more prolate-deformed ones form closed regions with many valence protons and neutrons.

For each LDM equation, nine fits were performed, one for each group of nuclei. In this way, nine sets of parameters were obtained, which minimize the RMS of each group of nuclei. The values of these parameters can be found in Tables 3, 5 and 7 from Ref. \cite{Hir11}. We present in Table \ref{tab2} the RMS obtained for all groups employing these nine sets of parameters.

\begin{table}[h]
\begin{center}
\caption {RMS (in keV) for each of the nine groups (columns), employing Eqs. (\ref{piet}), (\ref{wang}) and (\ref{roger}), respectively.}
\label{tab2}
\bigskip
\begin{tabular}{c|c|ccccccc|c}
\hline \hline
       & all  &	1	&2	&3	&4	&5	&6	&7	&semi-magic\\
\hline
LDM1&$2387$&$1313$&$1676$&$2063$&$1746$&$1053$&$870$&$746$&$2113$\\
LDM2&$2374$&$1254$&$1675$&$2069$&$1762$&$1021$&$838$&$656$&$2056$\\
LDM3&$2422$&$1183$&$1597$&$2151$&$1517$&$986$&$819$&$629$&$1967$\\
\hline\hline \end{tabular} \end{center}
\end{table}

These results show that for the three LDM models the masses of prolate deformed nuclei can be described with remarkable precision,  with an RMS smaller than $750$ keV, while the masses of spherical and semi-magic  nuclei are those worst described, with RMS larger than $2000$ keV. It completely challenged the authors preconception that the LDM was best suited to describe spherical nuclei, and that deformation effects were a crucial element necessary to improve the LDM description of nuclear binding energies. As can be read from Table \ref{tab2}, deformed regions are very well adjusted by the three LDM formulas, and the regions around shell closures are those which represent a challenge.

\begin{table}[h]
\begin{center}
\caption{Number of nuclei with even $N$-even $Z$ ($N_{nucl}^{ee}$), even $N$-odd $Z$ ($N_{nucl}^{eo}$), odd $N$-even $Z$ ($N_{nucl}^{oe}$) and odd $N$-odd $Z$ ($N_{nucl}^{oo}$).}
\label{eo}
\begin{tabular}{c||c|ccccccc|c}
\hline \hline
group & all  &	1	&2	&3	&4	&5	&6	&7	&semi-magic\\
\hline
$N_{nucl}^{ee}$  &576 & 62	& 35  &154	       &53       & 80 & 104  & 88   &95\\
$N_{nucl}^{eo}$  &534 & 67	& 74  &67	       &79       & 73 & 89  & 85   &40\\
$N_{nucl}^{oe}$  &536 & 69	& 60  &80	       &62       & 78 & 89  & 98   &50\\
$N_{nucl}^{oo}$  &503 & 60	& 83  &31	       &78       & 76 & 82  & 93  &0\\
\hline\hline \end{tabular} \end{center}
\end{table}

Given that the description of pairing effects for nuclei with few valence nucleons would require more than a simple parametrization as the one employed here, we have removed odd - even effects from our analysis, 
separating
the nuclei in four additional sets, corresponding to even $N$-even $Z$ (ee), even $N$-odd $Z$ (eo), odd $N$-even $Z$ (oe) and odd $N$-odd $Z$ (oo). The number of nuclear species of each type in each set is indicated in Table \ref{eo}. 
We repeated
the fits for each ee, eo, oe and oo set in the nine regions, employing the LDM1 mass formula but without the pairing term (last term in Eq. \rf{piet}). We arrive to the RMS indicated in Table \ref{pairing}, which one more time indicate that prolate deformed nuclei in region 7 are far better described than spherical ones, regions 2, 3, 4 and semimagic. It shows that
the difficulty of the LDM mass formulas to describe spherical nuclei is not at all associated with any odd-even effect, or with the way the pairing contribution is parameterized.

\begin{table}[h]
\begin{center}
\caption {RMS (in keV) for each of the nine groups (columns) in each ee, eo, oe and oo region, employing Eq. (\ref{piet}) without
pairing.}
\label{pairing}
\bigskip
\begin{tabular}{c|c|ccccccc|c}
\hline \hline
       & all  &	1	&2	&3	&4	&5	&6	&7	&semi-magic\\
\hline
even $N$-even $Z$&$2418$&$1158$&$1443$&$2272$&$1529$&$1050$&$805$&$786$&$2194$\\
even $N$-odd $Z$&$2404$&$1387$&$1834$&$1698$&$1726$&$1036$&$867$&$712$&$1483$\\
odd $N$-even $Z$&$2360$&$1382$&$1514$&$1936$&$1638$&$990$&$880$&$749$&$1871$\\
odd $N$-odd $Z$&$2404$&$1430$&$1370$&$1595$&$1888$&$1072$&$913$&$703$&$-$\\
\hline\hline \end{tabular} \end{center}
\end{table}

In what follows we explore the ability of the algebraic extensions of the LDM 
to include shell effects in the description of nuclear masses.

\section{The Duflo-Zuker inspired mass models}

In this section two different mass models, both based on Duflo-Zuker ideas, are discussed in detail.
The master terms are introduced in two alternative way in order to describe the shell effects in mass models, and their correlation with the nuclear deformation is analyzed.

\subsection{DZ1}

The macroscopic sector of the simplest version of the DZ mass model, contains six terms leading asymptotically to a LD form \cite{Men10} (see section C below)
\begin{equation}
E_{DZ1} = a_1(M^{HO} +S)-a_2\frac{M^{HO}}{\rho}-a_3 V_C -a_4 V_T +a_5 V_{TS} + a_6 V_P.
\label{dz1}
\end{equation}

The master $M^{HO}$ term is
\begin{equation}
M^{HO} =\frac{1}{2\rho}\left[\left(\sum_{p}\frac{n_\nu+n_\pi}{\sqrt{D_p^{HO}}}\right)^2
+\left(\sum_{p}\frac{n_\nu-n_\pi}{\sqrt{D_{p}^{HO}}}\right)^2\right],
\label{master}
\end{equation}
where the sums run over all occupied proton and neutron orbitals up to the Fermi level,
$D_p^{HO}=(p+1)(p+2)$ is the degeneracy of the major harmonic-oscillator (HO) shell of principal quantum number $p$, and 
$n_\nu(p)$, $n_\pi(p)$ are the number of
neutrons and protons, respectively, in the HO shell $p$.

It is relevant to mention that this form describes the dominant contribution of the monopole part of the nuclear Hamiltonian.
In order to change the HO closures (at $N, Z = 8$, $20$, $40$, $70$, $\cdots$) into the observed
extruder-intruder (EI) ones at $N, Z = 14$, $28$, $50$, $82$ and $126$,
the $S$ operator proposed by Duflo, given in Eq. (18) of Ref. \cite{Men10}, is employed.

The scaling factor is
\be
\rho= A^{1/3}\left[1 -\left(\frac{T}{A}\right)^2\right]^2.
\label{9}\ee

The Coulomb, asymmetry and surface asymmetry terms are \cite{Men10}, respectively:
\br
V_C &=&\frac{-Z(Z -1)+ 0.76[Z(Z -1)^{2/3}]}{r_c},\hspace{1cm}r_c= A^{1/3}\left[1 -\left(\frac{T}{A}\right)^2\right]\nn\\
V_T&=&\frac {4 T(T+1)}{A^{2/3}\rho},\nn\\
V_{TS}&=&\frac {4 T(T+1)}{A^{2/3}\rho^2}-\frac {4 T(T-\frac{1}{2})}{A\rho^4}.
\label{10}\er
The pairing term $V_P$ is the same employed in LDM2.

\subsection{DZ2}

As there exists some uncertainty in the parameterization of the monopole part of the nuclear Hamiltonian \cite{Men10}, in
Ref. \cite{Hir10} a modified master term $M$ was proposed, which directly includes the EI shell closures. It is built
 using an expression similar to \rf{master}, but with the index $p$ now referring to the EI major shell with degeneracies $D_p\rightarrow D_p^{HO}+2$.
\begin{equation}
E_{DZ2} = a_1 MA - a_2\frac{MA}{\rho}-a_3V_C -a_4V_T +a_5V_{TS} + a_6V_P,
\label{dz2}
\end{equation}
with the new master term:
\begin{equation}
MA=\frac{1}{\rho}\left(e_{1\nu}^2+e_{1\pi}^2\right),\hspace{1cm}
e_{1\nu}=\sum_{p_\nu}\frac{n_\nu}{\sqrt{D_{p_\nu}}},\hspace{1cm}
e_{1\pi}=\sum_{p_\pi}\frac{n_\pi}{\sqrt{D_{p_\pi}}},
\label{master2}
\ee
where $D_{p_{\nu,\pi}}=(p_{\nu,\pi}+1)(p_{\nu,\pi}+2)+2$ contains the $HO\rightarrow EI$ information and 
$n_\nu(p_{\nu})$, $n_\pi(p_{\pi})$ are the number of
neutrons and protons in their respective EI shell.

\subsection{The master term}

The master term $MA$ contains the
microscopic information about shell closures.
We show in Fig. \ref{fig2} its behavior as a function of $A$. 
It scales as a linear function in $A$. Therefore, we can interpreter this term as a {\it volume} one and, consequently, the term $MA/\rho$ can be associated to a {\it surface} one.
This allows us to consider the macroscopic sector of the DZ mass formula as an extension of the LDM.

\begin{figure}[h]
\vspace{-8.5cm}
\includegraphics[width=19.0cm]{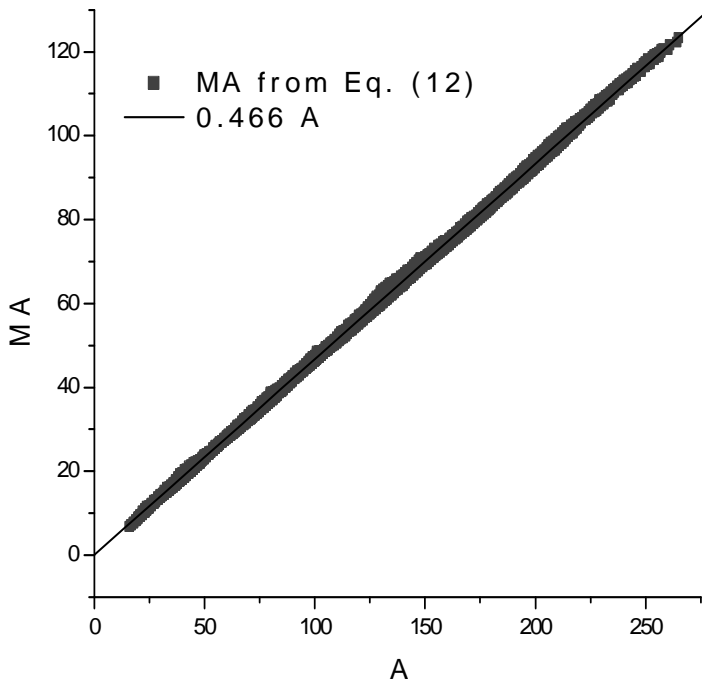}
\vspace{-7.5cm}
\caption{Master term from DZ model as function of $A$: gray squares indicate the results calculated with Eq. \rf{master2} and solid black line is 
a linear fit of them.}
\label{fig2}
\end{figure}

\begin{figure}[h]
\vspace{-3.5cm}
\includegraphics[width=12.0cm]{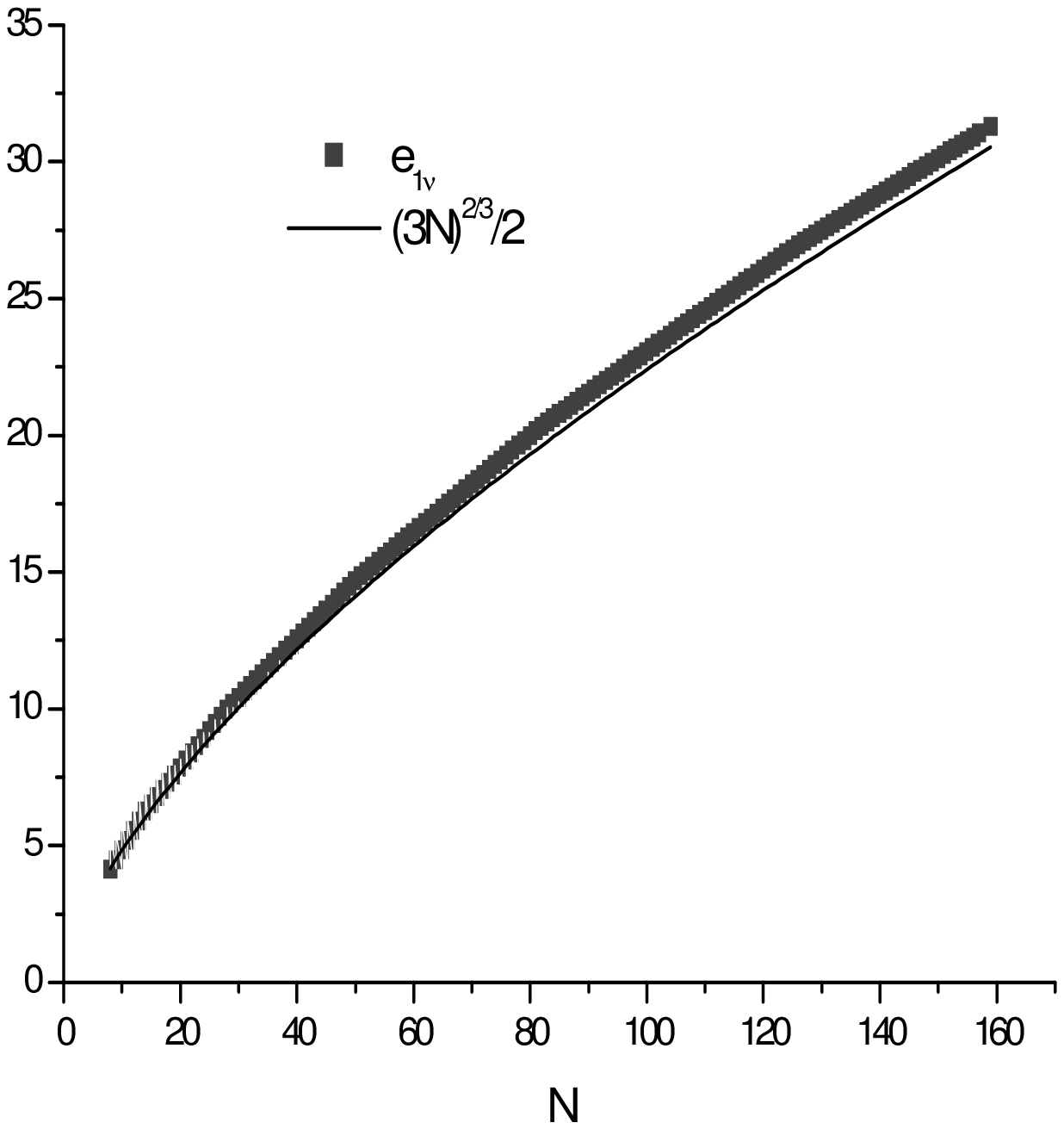}
\vspace{-5.0cm}
\caption{Dependence of $e_{1\nu}$ with $N$.}
\label{fig3}
\end{figure}

The building blocks of the master term $MA$ in Eq. (\ref{master2}) are $e_{1\nu}$ and $e_{1\pi}$.
In Fig. \ref{fig3} the behavior of $e_{1\nu}$ as a function of $N$ is presented. 
The thin black line represents the asymptotic form $e_{1\nu}\approx \frac{(3N)^{2/3}}{2}$ \cite{Hir10}.
The plot confirms that $e_{1\nu}$ scales as $N^{2/3}$, with minor differences for small $N$. To visualize this deviation,
the difference between $e_{1\nu}$ and this asymptotic behavior is 
presented in Fig. \ref{fig4} (black triangles).
The microscopic elements introduced through $e_{1\nu}$ are made evident in this plot. When the asymptotic behavior is removed, shell effects emerge, with well defined peaks at shell closures. However, there is a remnant continuous increase with N  which has not a linear dependence.
The description of the asymptotic form
of $e_{1\nu}$ can be improved by including terms with different powers of $N^{1/3}$.  The differences between $e_{1\nu}$ and these asymptotic forms are displayed with 
different symbols in Fig. \ref{fig4}. From these results we conclude that a better description can be reached if we adopt the new asymptotic form
\be
e_{1\nu,asym}=-0.90892 + 0.54259 N^{1/3} + 0.98851 N^{2/3} + 0.0018 N.
\label{asym}\ee
\begin{figure}[h!]
\vspace{-4.5cm}
\includegraphics[width=15.0cm]{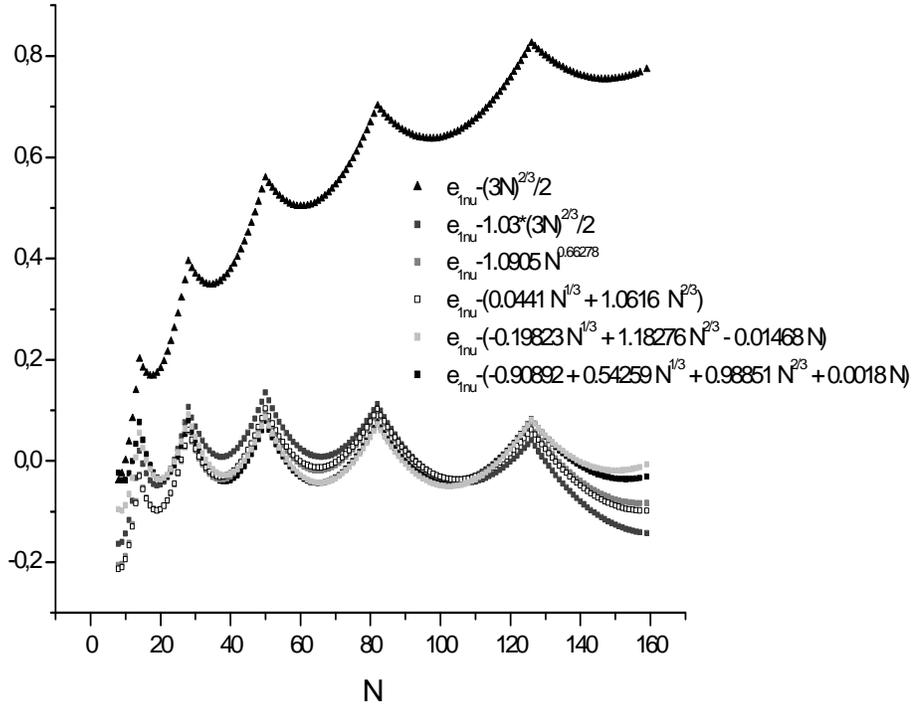}
\vspace{-4.5cm}
\caption{Difference between $e_{1\nu}$ and different approximations to describe its asymptotic behavior as a function of $N$.}
\label{fig4}
\end{figure}
A similar analysis was performed around Fig. 3 of Ref. \cite{Men10}.

In the construction of the master term $MA$, Eq. \rf{master2}, we 
employ $e_{1\nu}^2$. It is compared with its asymptotic form $e_{1\nu,asym}^2$ in Fig. \ref{fig5}.
The advantages of using the Eq. \rf{asym} instead of the %old
simpler $\frac{(3N)^{2/3}}{2}$ expression 
can be clearly observed: we obtain a global leveling besides the good description of closed shell effects.

\begin{figure}[h]
\vspace{-5.5cm}
\includegraphics[width=15.0cm]{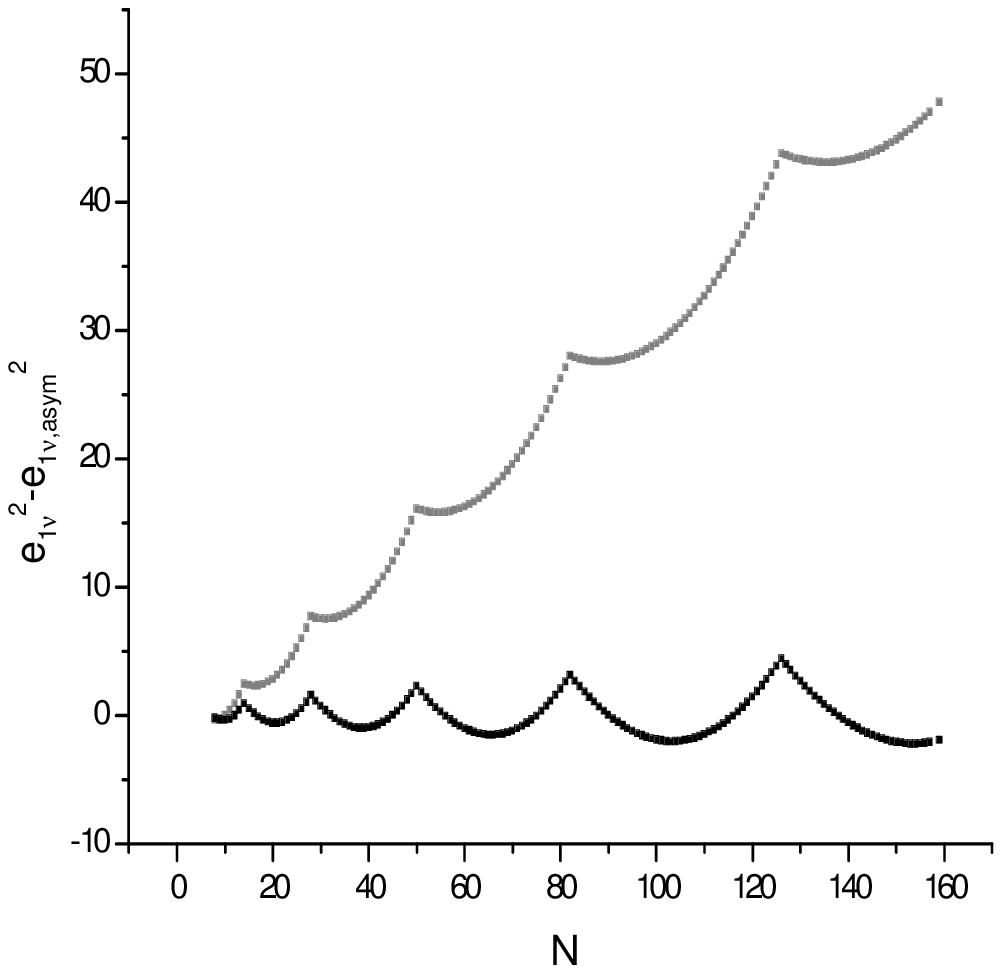}
\vspace{-6cm}
\caption{Difference between $e_{1\nu}^2$ and its asymptotic behavior for the approximation $e_{1\nu}\approx\frac{(3N)^{2/3}}{2}$ (gray line) and for Eq. \rf{asym} (black line).}
\label{fig5}
\end{figure}

Thus the master term $MA$ has the asymptotic behavior described by
\begin{equation}
MA_{asym}=\frac{1}{\rho}\left(e_{1\nu,asym}^2+e_{1\pi,asym}^2\right).
\label{master3asym}
\ee

\subsection{The fits}

Employing the two versions of the macroscopic sector of the DZ mass model, Eqs. \rf{dz1} and \rf{dz2}, we have fitted the nuclear masses in the nine regions described above.

\begin{table}[h]
\begin{center}
\caption {RMS (in keV) for each of the nine groups, using Eqs. (\ref{dz1}) and (\ref{dz2}).}
\label{tab3}
\bigskip
\begin{tabular}{c|c|ccccccc|c}
\hline \hline
       & all  &	1	&2	&3	&4	&5	&6	&7	&semi-magic\\
\hline
DZ1&$2852$&$994$&$1969$&$1557$&$1392$&$2237$&$2562$&$1529$&$1392$\\
DZ2&$3443$&$1425$&$1544$&$2167$&$1717$&$1729$&$2047$&$2107$&$1973$\\
\hline\hline \end{tabular} \end{center}
\end{table}

The fitting procedure in the different deformation regions leads to the RMS
displayed in
Table \ref{tab3}. These results show that the ability of both models to describe masses 
of nuclei in spherical, prolate and semi-magic groups are now comparable. The global RMS are larger than those obtained with the LDM formulas.
It is hard to find any correlation between the RMS and the regions with different deformations.
This fact is also reflected in the values of the fitted parameters, presented in Table \ref{tab4} and \ref{tab5} of the Appendix, because some of them (mainly $a_5$ and $a_6$) vary noticeably from one region to the other.

\section{Adding shell corrections to the LDM}

\subsection{LDM1 plus DZ}

It is possible to combine in a single mass formula both the ability of the LDM to describe the masses of deformed nuclei, and of DZ to include shell effects. To this end, we will work with a LDM, adding the shell effects present in the DZ master term, but with their asymptotic behavior removed, because it is basically contained in the volume and surface terms of the LDM. We have selected as the starting point the LDM1, which has a global RMS smaller than those of DZ1 and DZ2.

Based in our previous analysis, the shell effects will be introduced by a "volume" and "surface" shell corrections defined as
$MA-MA_{asym}$ and $(MA-MA_{asym})/\rho$, respectively. They are plotted in Fig. \ref{fig6} as a function of $A$.

\begin{figure}[h]
\vspace{-5cm}
\begin{center}$
\begin{array}{cc}
\hspace{-4cm}\includegraphics[width=15.1cm]{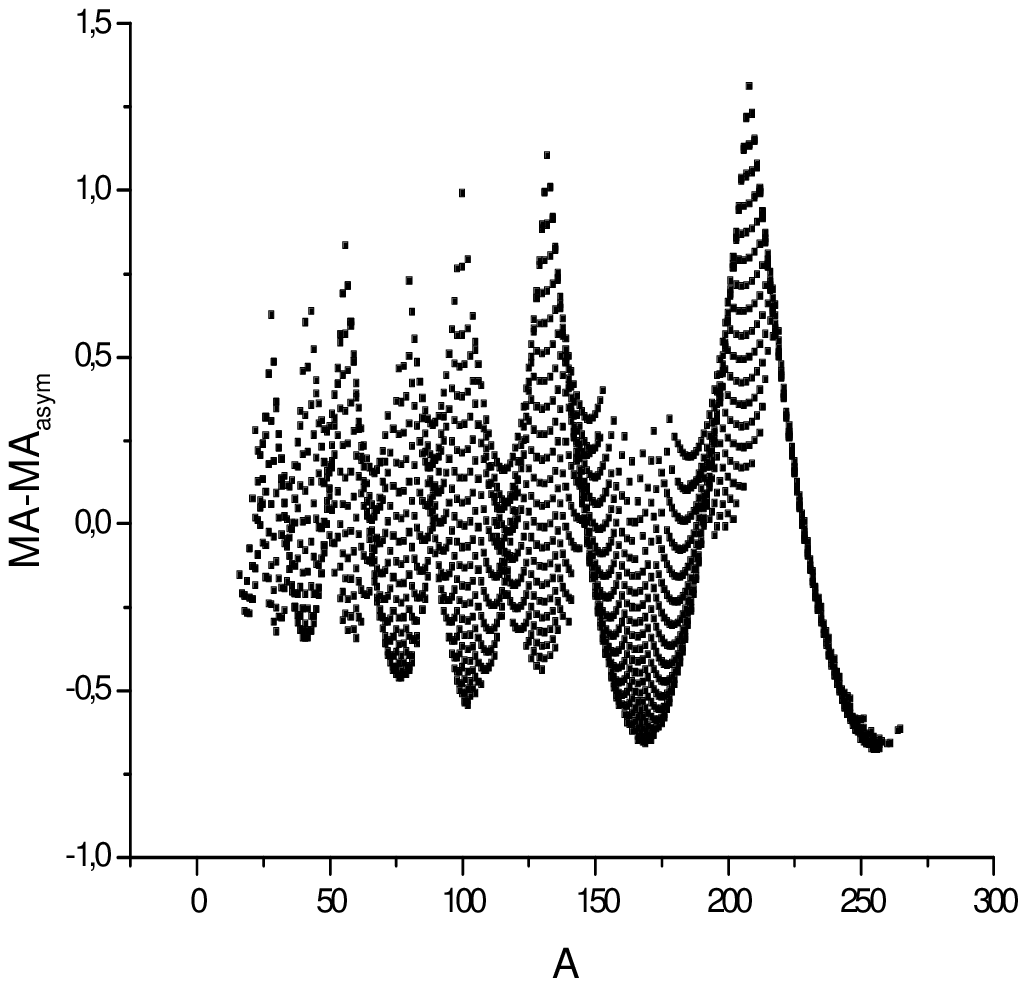} &
\hspace{-6cm}\includegraphics[width=15cm]{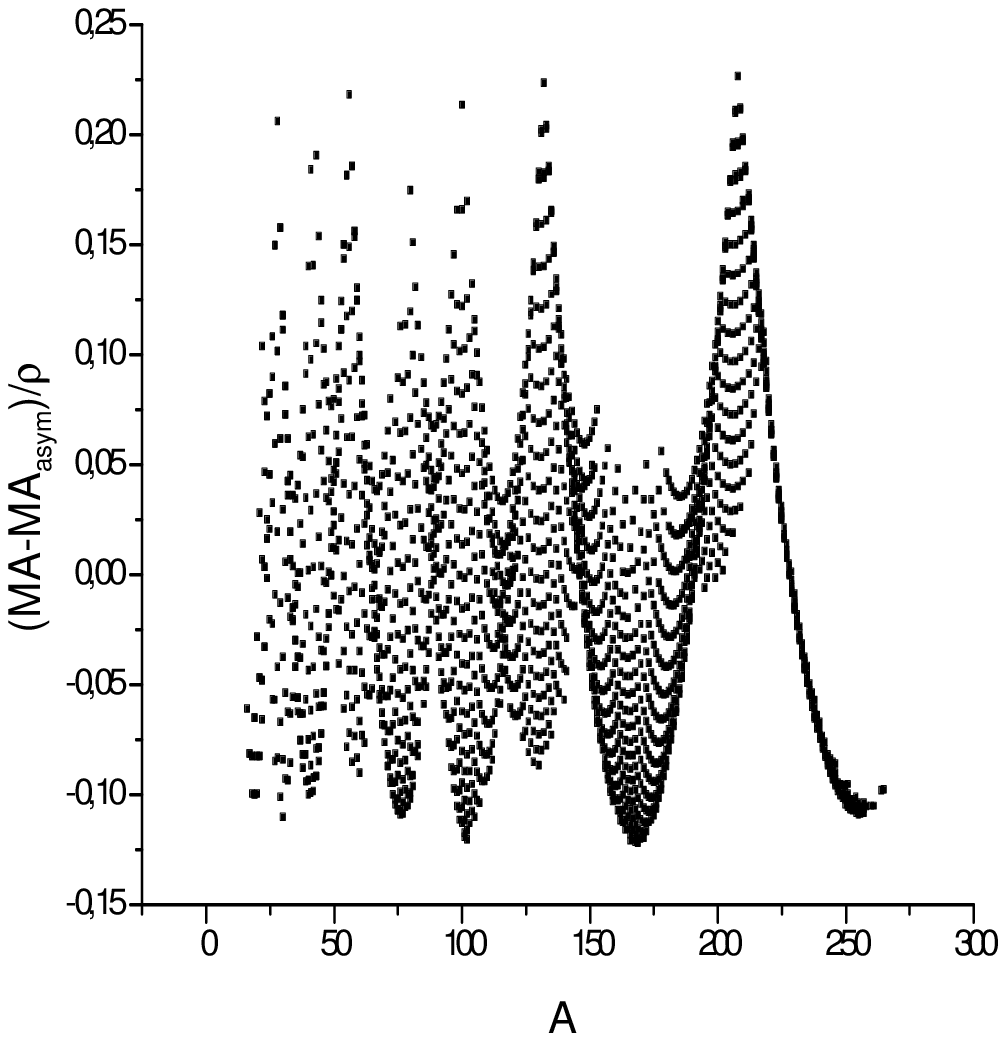}
\end{array}$
\end{center}
\vspace{-7cm}
\caption{Volume ($MA-MA_{asym}$) and surface ($(MA-MA_{asym})/\rho$) terms as function of $A$.}
\label{fig6}
%\vspace{4cm}
\end{figure}

They resemble the shell effects not included in the LDM.
To help the visualization of the shell effects introduced by these differences,
we show in Fig. \ref{fig7} the same terms as function of $N$, $Z$ and $A$
only for semi-magic nuclei, where we can observe one more time the 
clear peaks at the magic numbers.

\begin{figure}[h]
\vspace{-6.5cm}
\begin{center}$
\begin{array}{ccc}
\hspace{-5cm}\includegraphics[width=15cm]{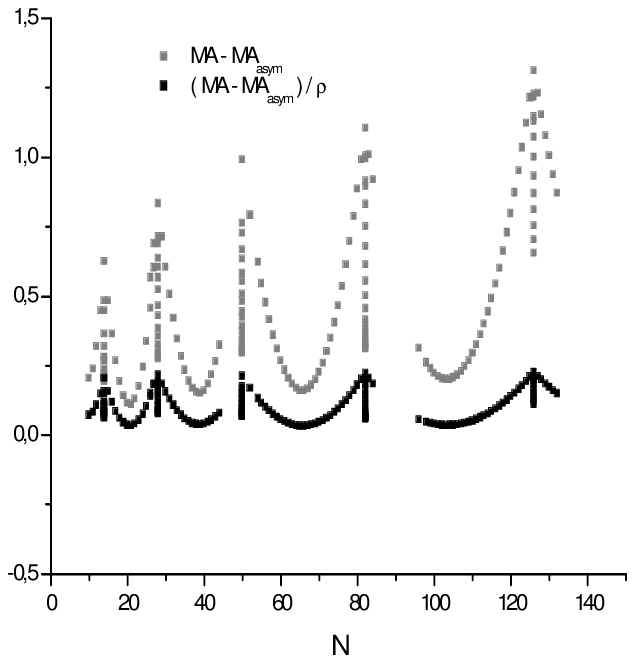} &
\hspace{-10cm}\includegraphics[width=15cm]{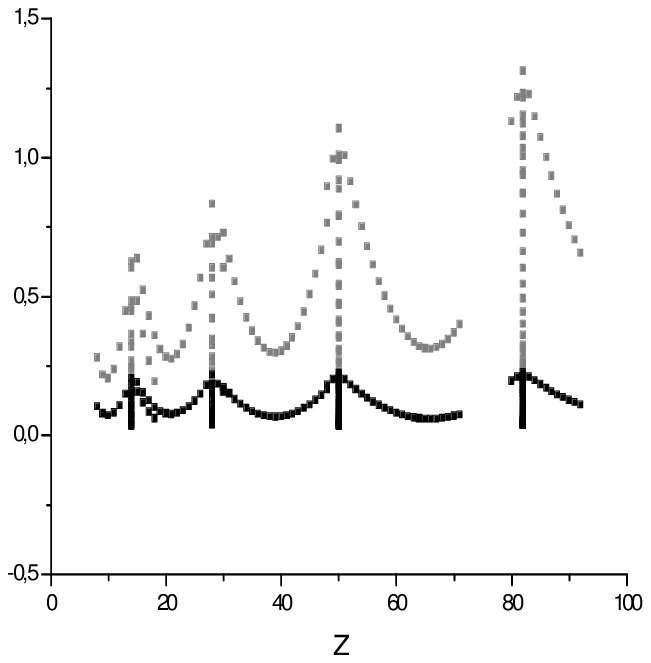} &
\hspace{-9cm}\includegraphics[width=15cm]{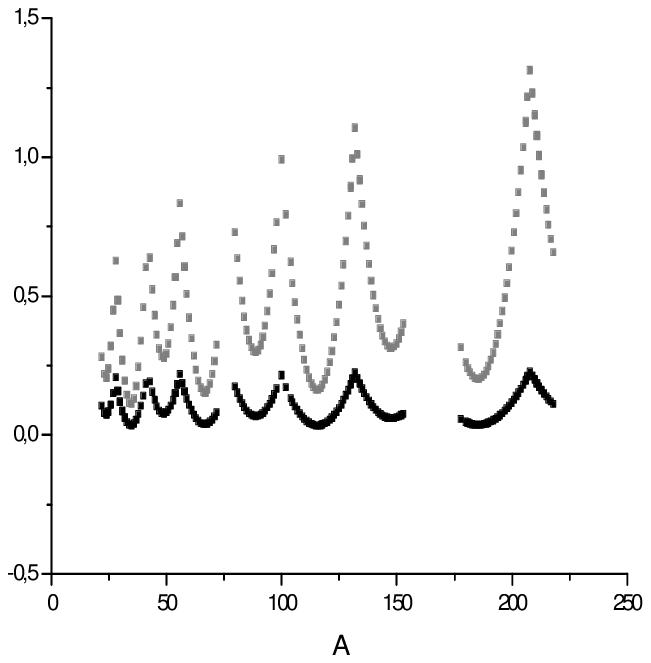}
\end{array}$
\end{center}
\vspace{-8cm}
\caption{Volume ($MA-MA_{asym}$) and surface ($(MA-MA_{asym})/\rho$) terms for semi-magic nuclei.}
\label{fig7}
\end{figure}

As explained above, we have added the shell effects to the LDM1 to construct a new mass formula:
\be
E_{LDM+DZ}=LDM1+a_{vol}(MA-MA_{asym})+a_{surf}(MA-MA_{asym})/\rho.
\label{newmass}\ee

\subsection{Comparison with other estimations of shell effects}

Inspired in the F-spin symmetry of the IBM's \cite{Die07,Die09}, shell effects have been recently introduced in extensions of the LDM by adding two terms, linear and quadratic in the number of valence protons, $z_v$, and neutrons, $n_v$ \cite{Die07,Men08,Die09}.
The proposed expression for the nuclear binding energies is
\be
E_{LDM+val}=LDM1+b_1(n_v+z_v)+b_2(n_v+z_v)^2.
\label{die}\ee

\subsection{The fits}

Repeating the analysis for the same nine groups, we obtain the results shown in Tables \ref{tab8} for the RMS 
employing the two models.

\begin{table}[h]
\begin{center}
\caption {RMS (in keV) for each of the nine groups, using Eqs. (\ref{newmass}) and (\ref{die}).}
\label{tab8}
\bigskip
\begin{tabular}{c|c|ccccccc|c}
\hline \hline
       & all  &	1	&2	&3	&4	&5	&6	&7	&semi-magic\\
\hline
LDM+DZ&$1407$&$668$&$907$&$1026$&$755$&$784$&$791$&$647$&$952$\\
LDM+val&$1075$&$796$&$981$&$1006$&$828$&$711$&$836$&$615$&$1037$\\
\hline\hline \end{tabular} \end{center}
\end{table}

As expected, the improvements in the inclusion of shell effects reduce the global RMS from $2387$ keV to $1407$ ($1075$) keV when the LDM+DZ (LDM+val) model is used. The results still show 
a visible tendency to describe better the deformed than spherical nuclei. Besides, the results obtained with both formulas look very similar, with
a smaller global RMS in the valence model and some advantage of the LDM+DZ model to describe semi-magic nuclei.

The parameters set obtained in both fits are presented in the Appendix, in Tables \ref{tab9} and \ref{tab10}.
The relative stability of the parameters $b_1$ and $b_2$ in the LDM+val model, shown in the last two columns of Table  \ref{tab10}, is behind the comparatively
small global RMS. On the other hand, the shell surface and volume coefficients $a_{surf}$ and $a_{vol}$ of the LDM+DZ model listed in the last two columns of Table  \ref{tab9} vary both in magnitude and in sign from one deformation region to another. Being a limitation for a good global fit, it offers at the same time the opportunity to relate these parameters with the deformation, a challenge which is left for future work.

Finally, to complete the comparison, we exhibit in Fig. \ref{fig8} the difference between experimental data ($B_{exp}$) and theoretical results ($B_{th}$) for both models. The two models which include shell effects display a clear reduction in the differences as compared with the Liquid Drop Model LDM1. As decades of work in nuclear masses have thought us, including these residual effects is a fairly non-trivial task.

\vspace{-4.5cm}
\begin{figure}[h]
\begin{center}$
\begin{array}{ccc}
\hspace{-2.7cm}\includegraphics[width=11cm]{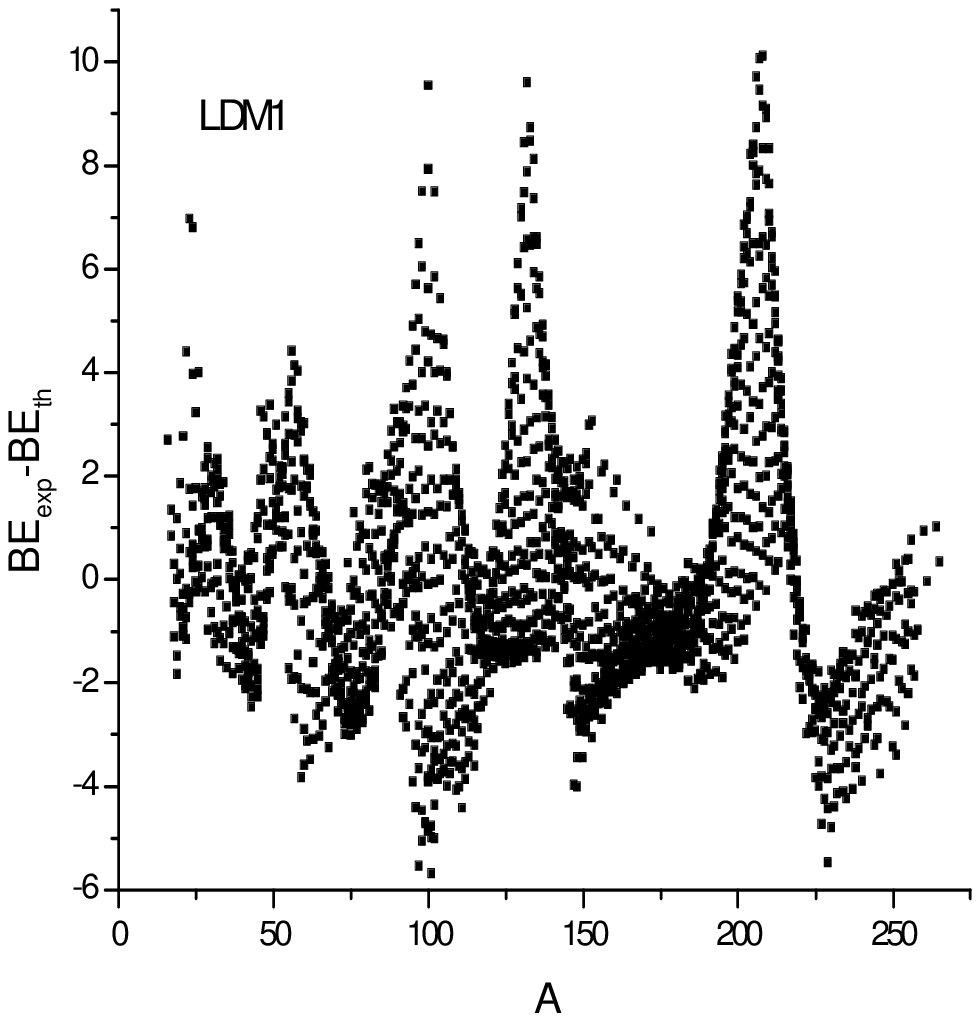} &
\hspace{-5.7cm}\includegraphics[width=11cm]{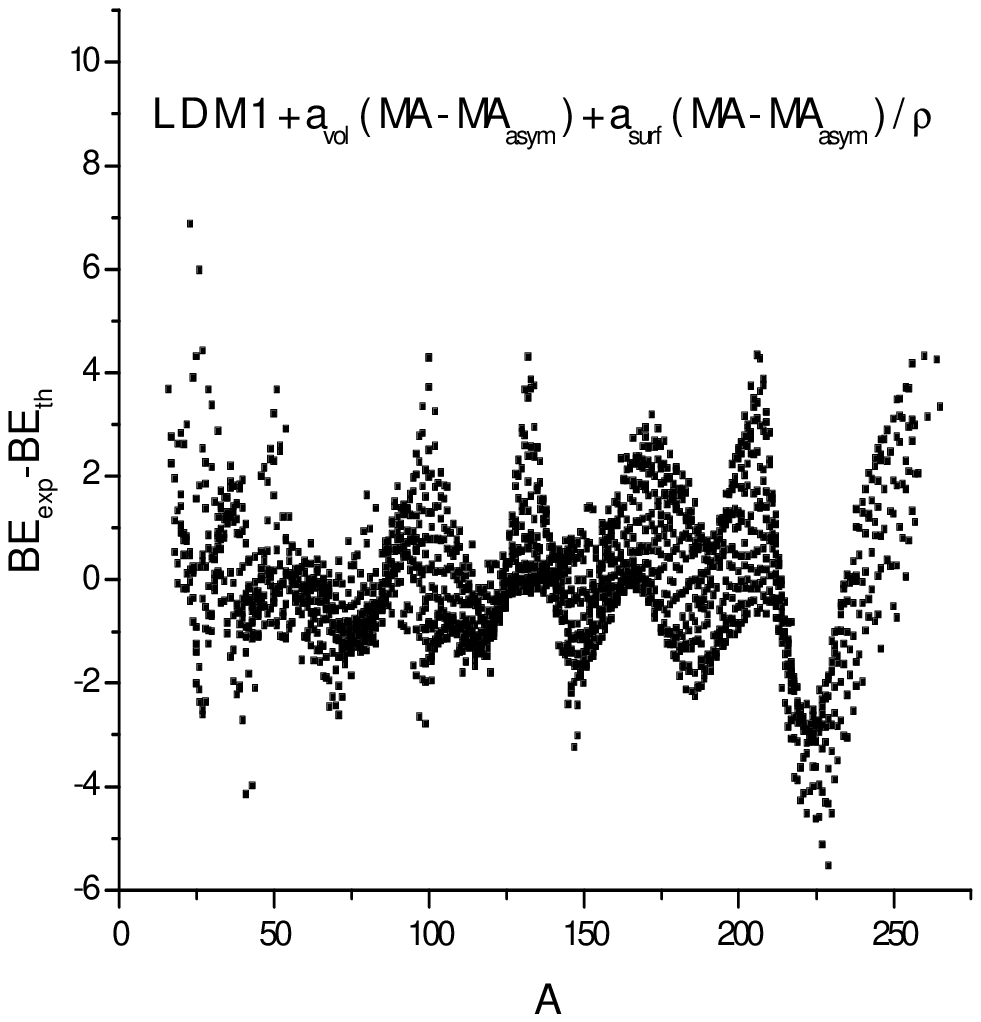} &
\hspace{-5.7cm}\includegraphics[width=11cm]{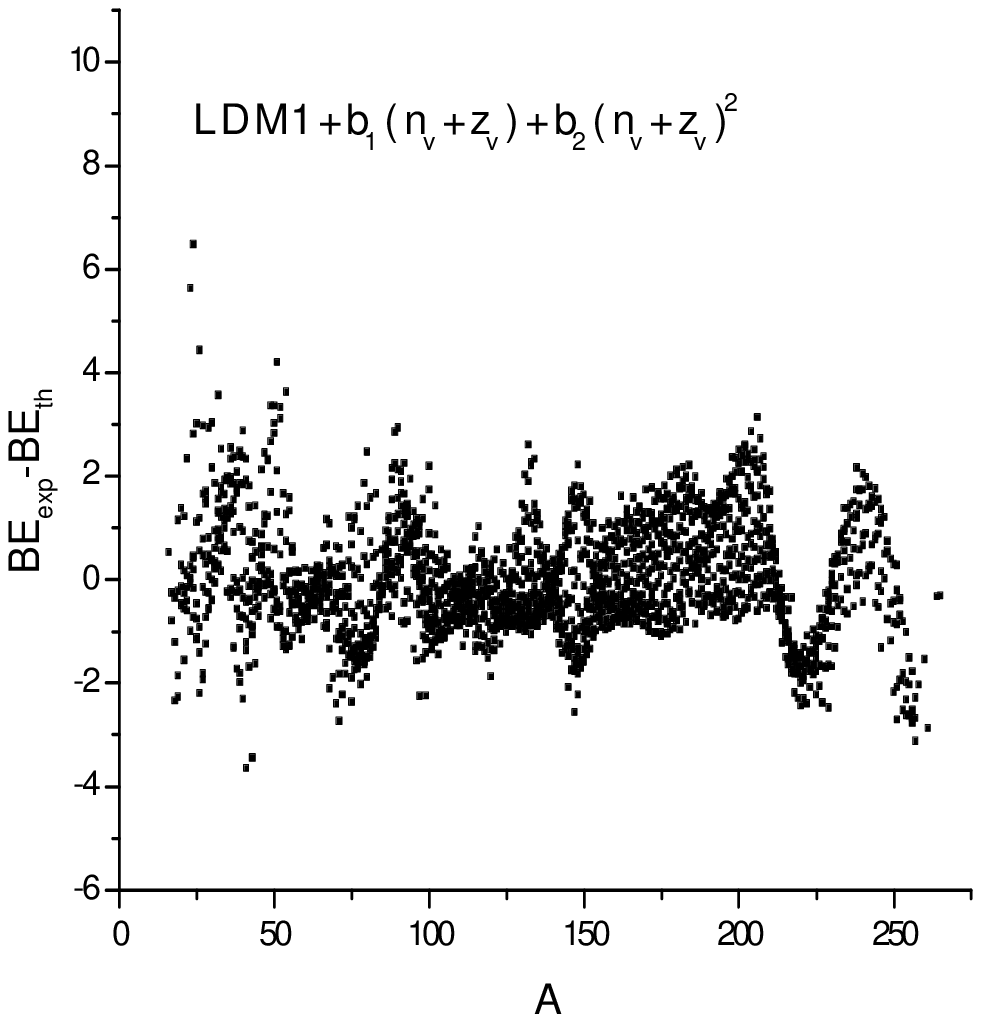}
\end{array}$
\end{center}
\vspace{-5cm}
\caption{Comparison between LDM1, LDM+DZ and LDM+val fits.}
\label{fig8}
\end{figure}

\section{Conclusions}

We have explored the ability of several Liquid Drop Models to describe nuclear binding energies in different deformation regions, and we have performed a similar analysis employing more elaborated models
which include shell effects through microscopic terms.
We selected for these microscopic corrections the Duflo-Zuker type models and those inspired in the Interacting Boson Model F-spin.

We have shown that the LDM is best suited to describe the masses of prolate deformed nuclei than of spherical ones, 
while these deformation effects are washed out employing the macroscopic sector of the DZ mass formulas.

We have paid special attention to the DZ master term, which
can be used to construct volume and surface terms which describe shell effects.
Adding these microscopic information to the LDM formula, we obtain an eight parameter 
 fit of nuclear masses, with a global RMS of $1407$ keV. The strong dependence of the parameters of the new shell volume and surface terms of the deformation regions could open the possibility to relate shell effects and deformation in an more elaborated way.

\acknowledgments

C.B. and A.M. are fellows of the CONICET, CCT La Plata (Argentina). 
JGH thanks A. Aprahamian, O. Civitarese, S. Pittel and P. van Isacker for their valuable comments.
This work was supported in part by the Agencia Nacional de Promoci\'on Cient\'{\i}fica y Tecnol\'ogica through PICT-2007-00861, Conacyt, M\'exico, and DGAPA, UNAM.

\newpage

\appendix{\bf {Appendix: Parameters of the fits}}

\bigskip

In this appendix we present the parameters found to provide the best fits of the nuclear masses in the nine regions for the mass equations DZ1 (Eq. \rf{dz1}), DZ2 (Eq. \rf{dz2}),  LDM1+DZ (Eq. \rf{newmass}) and LDM1+val (Eq. \rf{die}).
\begin{table}[h]
\begin{center}
\caption {Sets of parameters (in keV) which minimize the RMS for the nine groups of nuclei, employing DZ1, Eq. \rf{dz1}.}
\label{tab4}
\bigskip
\begin{tabular}{c|cccccc}
\hline \hline
&$a_1$&$a_2$&$a_3$&$a_4$&$a_5$&$a_6$\\
\hline
set$_{all}$&$17.531$&$15.433$&$0.694$&$36.157$&$48.425$&$5.154$\\
set$_1$&$17.458$&$15.204$&$0.691$&$33.959$&$39.860$&$5.857$\\
set$_2$&$17.760$&$16.283$&$0.706$&$38.212$&$57.641$&$6.167$\\
set$_3$&$17.788$&$16.329$&$0.708$&$39.263$&$61.183$&$5.807$\\
set$_4$&$17.975$&$16.971$&$0.716$&$39.733$&$61.192$&$6.056$\\
set$_5$&$17.894$&$16.593$&$0.715$&$38.553$&$56.304$&$5.997$\\
set$_6$&$17.172$&$14.375$&$0.667$&$35.674$&$47.421$&$4.684$\\
set$_7$&$17.379$&$15.037$&$0.682$&$33.507$&$38.541$&$5.206$\\
set$_{semi}$&$17.993$&$17.054$&$0.719$&$38.218$&$54.671$&$7.428$\\
\hline\hline \end{tabular} \end{center}
\end{table}

\begin{table}[h]
\begin{center}
\caption {Sets of parameters (in keV) which minimize the RMS for the nine groups of nuclei, employing DZ2, Eq. \rf{dz2}.}
\label{tab5}
\bigskip
\begin{tabular}{c|cccccc}
\hline \hline
&$a_1$&$a_2$&$a_3$&$a_4$&$a_5$&$a_6$\\
\hline
set$_{all}$&$15.707$&$16.155$&$0.611$&$29.800$&$34.307$&$7.337$\\
set$_1$&$15.798$&$16.383$&$0.619$&$29.892$&$33.626$&$6.882$\\
set$_2$&$15.801$&$16.357$&$0.620$&$30.997$&$37.886$&$3.960$\\
set$_3$&$15.535$&$15.684$&$0.599$&$31.938$&$45.543$&$6.855$\\
set$_4$&$16.248$&$17.924$&$0.646$&$31.604$&$36.711$&$8.869$\\
set$_5$&$16.116$&$17.509$&$0.635$&$32.334$&$42.394$&$6.581$\\
set$_6$&$15.685$&$16.315$&$0.598$&$31.551$&$42.802$&$6.118$\\
set$_7$&$15.381$&$15.076$&$0.589$&$25.663$&$18.329$&$7.087$\\
set$_{semi}$&$16.204$&$18.026$&$0.638$&$32.570$&$43.255$&$11.128$\\
\hline\hline \end{tabular} \end{center}
\end{table}

\begin{table}[h]
\begin{center}
\caption{Sets of parameters (in keV) which minimize the RMS for the nine groups of nuclei, using LDM1+DZ, Eq. \rf{newmass}.}
\vspace{.5cm}
\label{tab9}
\bigskip
\begin{tabular}{c|ccccccrr}
\hline \hline
&$a_v$&$a_s$&$a_c$&$a_p$&$S_v$&$y$&$a_{vol}$&$a_{surf}$\\
\hline
set$_{all}$&$15.817$&$18.422$&$0.705$&$5.917$&$29.740$&$2.298$&$2.700$&$12.526$\\
set$_1$&$15.727$&$18.142$&$0.700$&$5.893$&$28.721$&$2.090$&$17.792$&$-47.797$\\
set$_2$&$15.889$&$18.646$&$0.710$&$4.980$&$32.120$&$2.937$&$18.350$&$-48.154$\\
set$_3$&$15.805$&$18.409$&$0.703$&$5.712$&$33.574$&$3.352$&$20.536$&$-52.987$\\
set$_4$&$16.026$&$19.043$&$0.719$&$5.624$&$32.050$&$2.577$&$26.133$&$-80.211$\\
set$_5$&$15.911$&$18.717$&$0.711$&$5.731$&$31.079$&$2.532$&$1.962$&$13.575$\\
set$_6$&$15.703$&$18.103$&$0.696$&$5.297$&$29.991$&$2.368$&$-13.570$&$67.354$\\
set$_7$&$15.515$&$17.641$&$0.682$&$5.267$&$27.509$&$1.894$&$-10.468$&$47.078$\\
set$_{semi}$&$15.992$&$19.008$&$0.717$&$7.551$&$31.819$&$2.616$&$22.776$&$-59.556$\\
\hline\hline \end{tabular} \end{center}
\end{table}

\begin{table}[h]
\begin{center}
\caption{Sets of parameters (in keV) which minimize the RMS for the nine groups of nuclei using LDM1+val, Eq. \rf{die}.}
\vspace{.5cm}
\label{tab10}
\bigskip
\begin{tabular}{c|cccccccc}
\hline \hline
&$a_v$&$a_s$&$a_c$&$a_p$&$S_v$&$y$&$b_1$&$b_2$\\
\hline
set$_{all}$&$15.887$&$18.232$&$0.712$&$5.651$&$31.242$&$2.565$&$0.771$&$-0.013$\\
set$_1$&$15.805$&$18.067$&$0.707$&$5.882$&$29.052$&$2.073$&$0.606$&$-0.008$\\
set$_2$&$15.870$&$18.038$&$0.715$&$4.555$&$29.963$&$2.224$&$0.848$&$-0.017$\\
set$_3$&$15.793$&$17.927$&$0.704$&$5.115$&$32.858$&$3.138$&$0.780$&$-0.014$\\
set$_4$&$16.075$&$18.789$&$0.725$&$5.086$&$31.544$&$2.401$&$0.730$&$-0.012$\\
set$_5$&$15.929$&$18.397$&$0.715$&$5.715$&$31.276$&$2.529$&$0.781$&$-0.017$\\
set$_6$&$15.872$&$18.456$&$0.708$&$5.795$&$30.722$&$2.435$&$0.380$&$-0.006$\\
set$_7$&$15.678$&$17.850$&$0.695$&$5.083$&$29.023$&$2.146$&$0.456$&$-0.009$\\
set$_{semi}$&$16.091$&$18.824$&$0.726$&$6.796$&$31.392$&$2.418$&$0.868$&$-0.017$\\
\hline\hline \end{tabular} \end{center}
\end{table}


\begin{thebibliography}{99}
\bibitem{Wang10} N. Wang, M. Liu and X. Wu, Phys. Rev. {\bf C81} (2010) 044322 .
\bibitem{Bet36} H.A. Bethe and R.F. Bacher, 
%{\it Nuclear Physics I. Stationary States of Nuclei}, 
Rev. Mod. Physics {\bf 8} (1936) 82-229.
\bibitem{Wik} http://en.wikipedia.org/wiki/Semi-empirical mass formula.
\bibitem{Fed79} P. Federman and S. Pittel, Phys. Rev. {\bf C20}, (1979) 820.
\bibitem{Bohr} A. Bohr and B.R. Mottelson, {\it Nuclear Structure} v. I, (World Scientific, Singapore, 1998).
\bibitem{Rol88} C.E.~Rolfs and W.S.~Rodney, {\it Cauldrons in the Cosmos} (University of Chicago Press, Chicago, 1988).
\bibitem{Bla06} Klaus Blaum, Phys. Rep. {\bf 425} (2006) 1.
\bibitem{Lunn03} D.~Lunney, J.M.~Pearson, C.~Thibault, Rev. Mod. Phys. {\bf 75} (2003) 1021.
\bibitem{Mol95} P.~M\"oller, J.R.~Nix, W.D.~Myers, W.J.~Swiatecki,
%{\it Nuclear ground-state masses and deformations.}  
At. Data Nucl. Data Tables {\bf 59} (1995) 185.
\bibitem{Wan10} N. Wang, Z. Liang, M. Liu and X. Wu, Phys. Rev. {\bf C 82} (2010) 044304.
\bibitem{Mye96} W.D. Myers and W.J. Swiatecki, Nucl. Phys. {\bf A 601} (1996) 141.
\bibitem{Pom03} K. Pomorski and J. Dudek, Phys. Rev. {\bf C67}, (2003) 044316.
\bibitem{Gor01} S.~Goriely, F.~Tondeur, J.M.~Pearson, 
%A Hartree-Fock nuclear mass table.  
Atom. Data Nucl. Data Tables {\bf 77} (2001) 311;
S.~Goriely, M.~Samyn, J.M.~Pearson, Phys. Rev. {\bf C75} (2007) 064312.
\bibitem{Gor09} S.~Goriely, N. Chamel, J.M.~Pearson, Phys. Rev. Lett. {\bf 102} (2009) 152503;
S.~Goriely, S. Hilaire, M. Girod, adn S. P\'eru, Phys. Rev. Lett. {\bf 102} (2009) 242501.
\bibitem{Duf94} J.~Duflo, Nucl. Phys. {\bf A576} (1994) 29.
\bibitem{Zuk94} A.P.~Zuker, Nucl. Phys. {\bf A576} (1994) 65.
\bibitem{Duf95} J.~Duflo and A.P.~Zuker, Phys. Rev. {\bf C52} R23 (1995) R23.
\bibitem{Men08} J. Mendoza-Temis, I. Morales, J. Barea, A. Frank, J.G. Hirsch, J.C. L\'opez-Vieyra, P. Van Isacker and V. Vel\'azquez, Nucl. Phys. {\bf A812} (2008) 28.
\bibitem{Men10} J. Mendoza-Temis, J.G. Hirsch and A.P. Zuker, Nucl. Phys. {\bf A843} (2010) 14.
\bibitem{Hir10} J.G. Hirsch and J. Mendoza-Temis,  J. Phys. G: Nucl. Part. Phys.  {\bf 37} (2010) 064029.
\bibitem{Die07} A.E.L. Dieperink, P. Van Isacker, Eur. Phys. J. {\bf A 32} (2007) 11.
\bibitem{Die09} A.E.L.~Dieperink, P.~Van~Isacker, Eur.\ Phys. J. {\bf A 42} (2009) 269.
\bibitem{Gan11} G. Gangopadhyay, Int. J. Mod. Phys. {\bf E 20} (2011) 179.
\bibitem{Hir11} J.G. Hirsch, C. Barbero and A. Mariano, arXiv:1108.0707; J. Phys.: Conference Series (in press).
\bibitem{Roger10} G. Royer, M. Guilbaud and A. Onillon, Nucl. Phys. {\bf A847} (2010) 24.
\bibitem{AME03} G.~Audi, A.H.~Wapstra, C.~Thibault, Nucl. Phys. {\bf A 729} (2003) 337.
\bibitem{Minuit} F. James, Minuit: Function Minimization and Error Analysis Reference Manual, Version 94.1, CERN (1994); http://wwwasdoc.web.cern.ch/wwwasdoc/minuit/minmain.html
\bibitem{Moll95} P. M\"oller, J. R. Nix, W. D. Myers, and W. J. Swiatecki, Atomic Data Nucl. Data Tables {\bf 59} (1995) 185.
\end{thebibliography}
\end{document}